\documentclass{PoS}

\title{Scaling behavior of chiral phase transition in two-flavor QCD with improved Wilson quarks at finite density}

\ShortTitle{Scaling behavior of chiral phase transition with improved Wilson quarks}

\author{\speaker{Shinji Ejiri},
        Y.~Nakagawa\\
        Graduate School of Science and Technology, Niigata University, Niigata 950-2181, Japan\\
        E-mail: \email{ejiri@muse.sc.niigata-u.ac.jp}
}
\author{S.~Aoki, K.~Kanaya, H.~Ohno, H.~Saito\\
        Graduate School of Pure and Applied Sciences, University of Tsukuba, Tsukuba, Ibaraki 305-8571, Japan
}
\author{T.~Hatsuda\\
Department of Physics, The University of Tokyo, Tokyo 113-0033, Japan
}
\author{Y.~Maezawa\\
Mathematical Physics Laboratory, RIKEN Nishina Center, Saitama 351-0198,
Japan
}
\author{T.~Umeda\\
Graduate School of Education, Hiroshima University, Hiroshima 739-8524, Japan
}
\author{(WHOT-QCD collaboration)}

\abstract{
We study scaling behavior of a chiral order parameter performing a simulation of two-flavor QCD with improved Wilson quarks. It has been shown that the scaling behavior of the chiral order parameter defined by a Ward-Takahashi identity agrees with the scaling function of the three-dimensional $O(4)$ spin model at zero chemical potential. We extend the scaling study to finite density QCD. Calculating derivatives of the chiral order parameter with respect to the chemical potential in two-flavor QCD, the scaling property of chiral phase transition is discussed in the low density region. 
We moreover calculate the curvature of the phase boundary of the chirl phase transition in the temperature and chemical potential plane assuming the $O(4)$ scaling relation.
}

\FullConference{The XXVIII International Symposium on Lattice Filed Theory\\
		June 14-19,2010\\
		Villasimius, Sardinia Italy}

\begin{document}

\section{Introduction}
\label{sec:intoro}

It is important to study the nature of QCD phase transition at high temperature and low density for understanding the evolution of the early universe and analyzing experimental data obtained by the heavy ion corrosions.
In order to extract unambiguous signals for the QCD phase transition 
from the heavy ion collisions, quantitative calculations 
directly from the first principles of QCD are indispensable. 
At present, the lattice QCD simulation is the only systematic method to do so.

Many important properties of finite temperature quark matter have been uncovered by lattice simulations. 
Recent developments of computational techniques enabled us to extend the study to small chemical potentials.
Most lattice QCD studies at finite temperature $(T)$ 
and chemical potential $(\mu_q)$ have been performed using staggered-type 
quark actions with the fourth-root trick of the quark determinant. 
Because the theoretical base for the fourth-root trick is not confirmed, it is indispensable to carry out simulations adopting different lattice quark actions to control and estimate systematic errors due to lattice discretization.
Another unsatisfactory point of the staggered-type quarks is that, it was very difficult until recently \cite{bnl-bie09} to confirm the scaling properties around the critical point, which is universal to the three-dimensional $O(4)$ spin model for 2-flavor QCD, as expected from the effective sigma model analysis
\footnote{Precisely speaking, the chiral transition of 2-flavor QCD with staggered-type quarks is universal to the $O(2)$ spin model, whose scaling properties are close to those of the O(4) model.}. 
This may suggest large lattice artifacts in the results with staggered-type quarks.

Several years ago, the CP-PACS Collaboration has studied finite-temperature QCD 
using the clover-improved Wilson quark action coupled with the RG-improved Iwasaki glue \cite{cppacs1,cppacs2}.
With 2-flavors of dynamical quarks, the phase structure, the transition temperature and the equation of state have been investigated.
In contrast to the case of staggered-type quarks, both with the standard Wilson quark action \cite{iwasaki} and with the clover-improved Wilson quark action \cite{cppacs1}, the subtracted chiral condensate shows the scaling behavior with the critical exponents and scaling function of the $O(4)$ spin model in a rather wide range of the parameter space near the chiral phase transition. 

\begin{figure}[t]
\begin{center}
\includegraphics[width=2.5in]{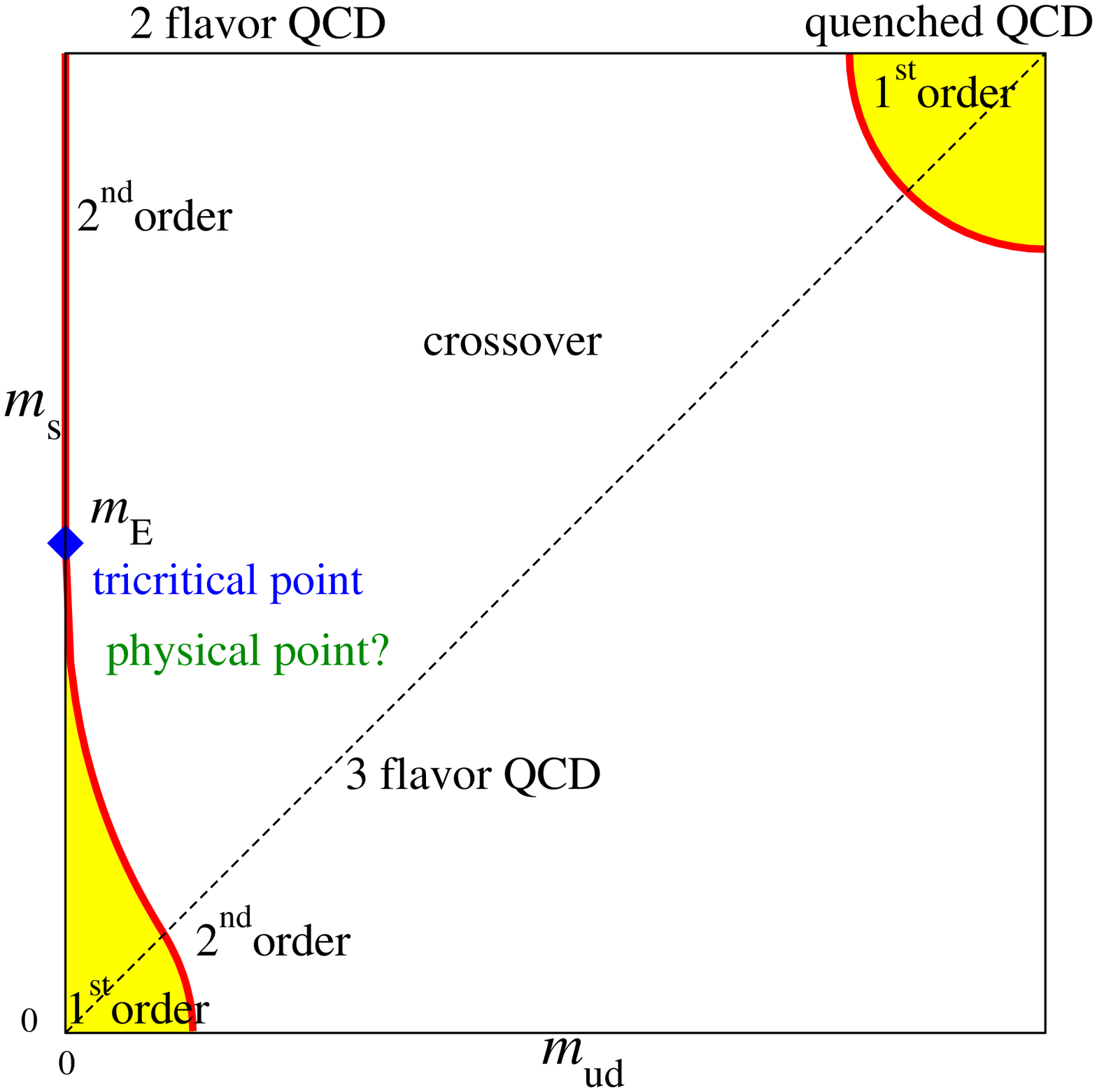}
\hskip 0.5cm
\includegraphics[width=2.5in]{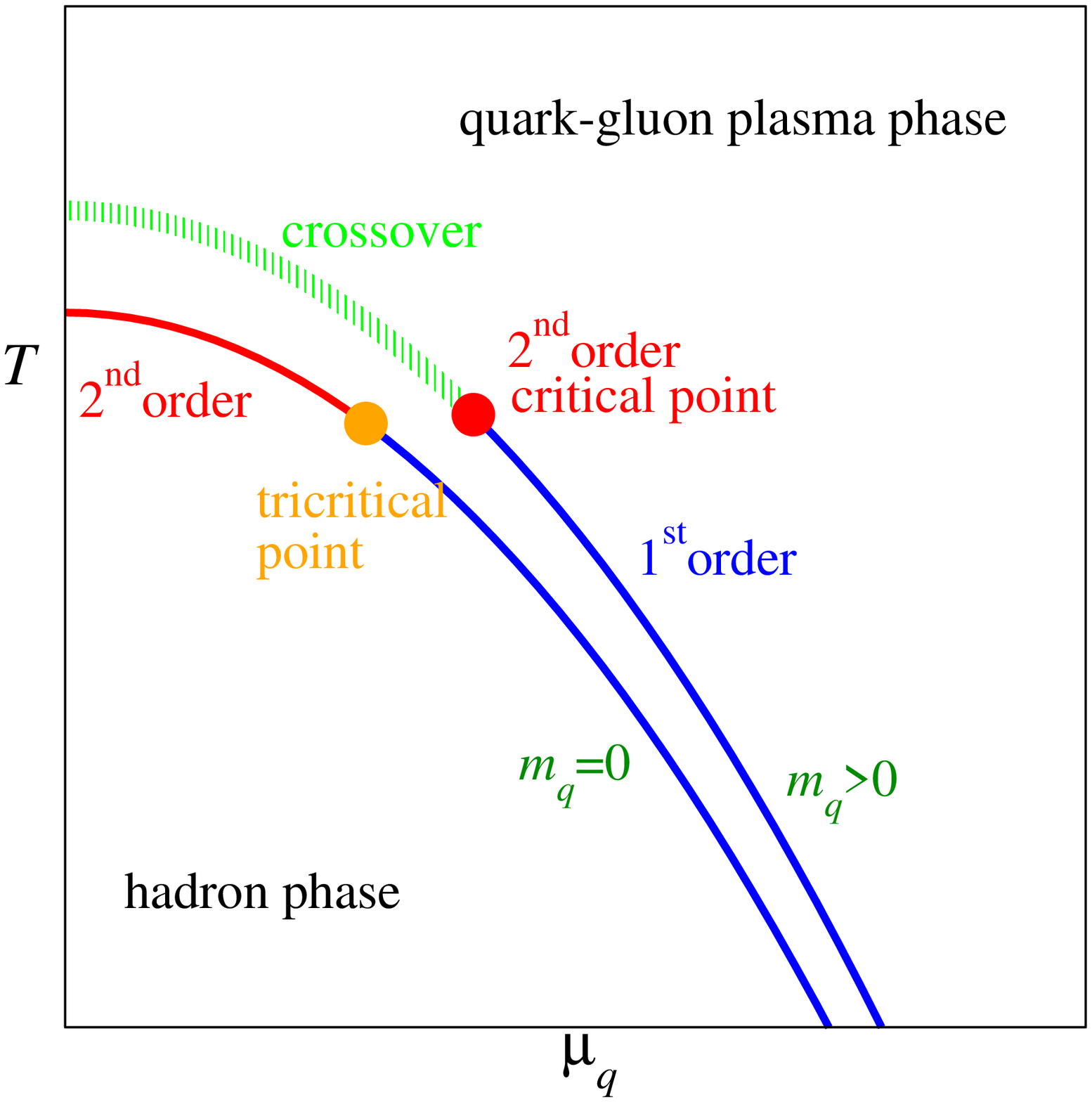}
\vskip -0.2cm
\caption{
Left: Mass-dependence of the nature of (2+1)-flavor QCD.
Right: Expected phase structure of 2-flavor QCD at finite temperature and density 
for $m_q=0$ and $m_q \neq 0$.
}
\label{fig1}
\end{center}
\vskip -0.3cm
\end{figure}

We would like to highlight scaling properties of the QCD phase transition by numerical simulations with dynamical Wilson quarks. 
As shown in Fig.~\ref{fig1} (left) for $\mu_q=0$, the phase transition is expected to be first order when the up and down quark masses $(m_{ud})$ and strange quark mass $(m_s)$ are sufficiently large or small, and becomes crossover in the intermediate region between them.
We expect a second order transition in the chiral limit of 2-flavor QCD, i.e. $ m_{ud} \equiv m_q =0, m_s= \infty $, and the nature of the transition changes as $m_s$ decreases.
It becomes first order below the trictitical point $(m_E)$.
When $\mu_q \neq 0$, the phase diagram in the $(T, \mu_q)$ plane is expected as Fig.~\ref{fig1} (right) for 2-flavor QCD with $m_q=0$ and $m_q \neq 0$. 
At low density, the phase transition is expected to be second order for $m_q=0$ and crossover for $m_q \neq 0$.
To confirm this phase structure, the scaling study in the crossover region is important. 
Recently, the scaling study with the physical strange mass has been done using a staggered-type quark action in Ref.~\cite{bnl-bie09}, and the scaling property including a chemical potential has been also discussed \cite{bnl-bie10}. 
Therefore, it is worth revisiting the scaling study using the Wilson-type quark action. 
In particular, we want to clarify the scaling property at finite $\mu_q$.

In this report, we study 2-flavor QCD, as a first step.
Previous results for $O(4)$ scaling at $\mu_q=0$ are summarized in Sec.~\ref{sec:scaling}. 
We then extend the discussion to finite density in Sec.~\ref{sec:dense}. 
The scaling properties are tested by numerical simulations in Sec.~\ref{sec:simu}. The conclusion is given in Sec.~\ref{sec:summary}.

\section{Scaling behavior of the chiral order parameter for 2-flavor QCD}
\label{sec:scaling}

The chiral phase transition in 2-flavor QCD at finite temperature is expected to have the same critical properties as the 3-dimensional $O(4)$ spin model. 
Hence, the scaling tests of the chiral order parameter 
$\langle \bar{\psi} \psi \rangle$ 
provide us with a useful way to study universality properties of the chiral phase transition for 2-flavor QCD. 

The order parameter of the $O(4)$ spin model is given by the magnetization $M$.
In the vicinity of the second order critical point, $M$ satisfies the following scaling relation:
\begin{eqnarray}
M/h^{1/\delta} = f(t/h^{1/y}),
\label{eq:o4}
\end{eqnarray}
where $h$ is the external magnetic field, $t$ is the reduced temperature which is defined by $t=(T-T_c|_{h=0})/T_c|_{h=0}$. The critical exponents have the values 
$1/y \equiv 1/(\beta \delta) = 0.537(7)$ and $1/ \delta = 0.2061(9)$,
and $f(x)$ is the scaling function.
In 2-flavor QCD, we identify $M \sim \langle \bar{\psi} \psi \rangle$, 
$h \sim 2m_q a$ and $t \sim \beta-\beta_{ct}$.
Here, $\beta_{ct}$ is the critical point in the chiral limit, and $a$ is the lattice spacing.
We compare the scaling functions of 2-flavor QCD and $O(4)$ spin model.

A careful treatment must be required because the chiral symmetry is explicitly broken for Wilson quarks at finite $a$. In Ref.~\cite{cppacs1,iwasaki}, it have been shown that the $O(4)$ scaling of Eq.~(\ref{eq:o4}) is well satisfied when one defines the quark mass $m_q a$ and the chiral order parameter $\langle \bar{\psi} \psi \rangle$ by Ward-Takahashi identities \cite{bochicchio}.
The quark mass is defined by the following correlation function for temporal direction:
\begin{eqnarray}
2m_q a = -m_{\rm PS} \left. \langle \bar{A_4}(t) \bar{P}(0) \rangle \right/
\langle \bar{P}(t) \bar{P}(0) \rangle,
\end{eqnarray}
where $P$ and $A_{\mu}$ are the pseudo-scalar and vector meson operators, respectively, 
$m_{\rm PS}$ is the pseudo-scalar meson mass, and the bar means the spatial average.
The chiral order parameter is given by the following equation:
\begin{eqnarray}
\langle \bar{\psi} \psi \rangle
= \frac{2 m_q a}{N_s^3 N_t} \sum_{x,x'} \langle P(x) P(x') \rangle
= \frac{2 m_q a (2K)^2 }{N_s^3 N_t} \left\langle 
{\rm tr} \left( D^{-1} \gamma_5 D^{-1} \gamma_5 \right) \right\rangle.
\end{eqnarray}
Here, $D$ is the quark matrix. 
The quark mass and the chiral condensate satisfy the Ward-Takahashi identity in the continuum limit:
\begin{eqnarray}
\langle \partial_{\mu} A_{\mu} (x) P(x') \rangle
-2m_q a \langle P(x) P(x') \rangle
=\delta(x- x') \langle \bar{\psi} \psi \rangle.
\end{eqnarray}

\begin{figure}[t]
\begin{center}
\includegraphics[width=2.8in]{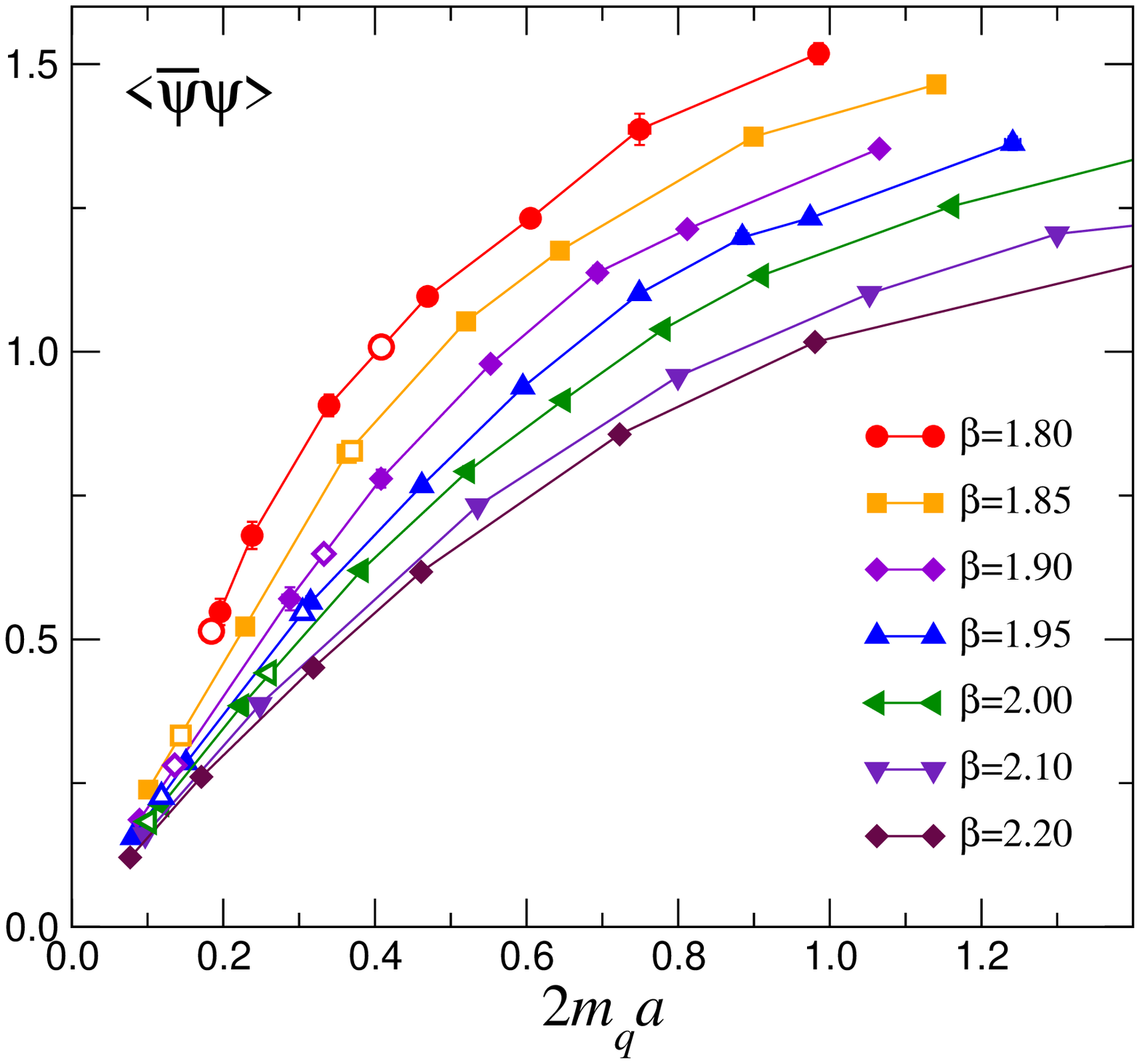}
\hskip 0.3cm
\includegraphics[width=2.8in]{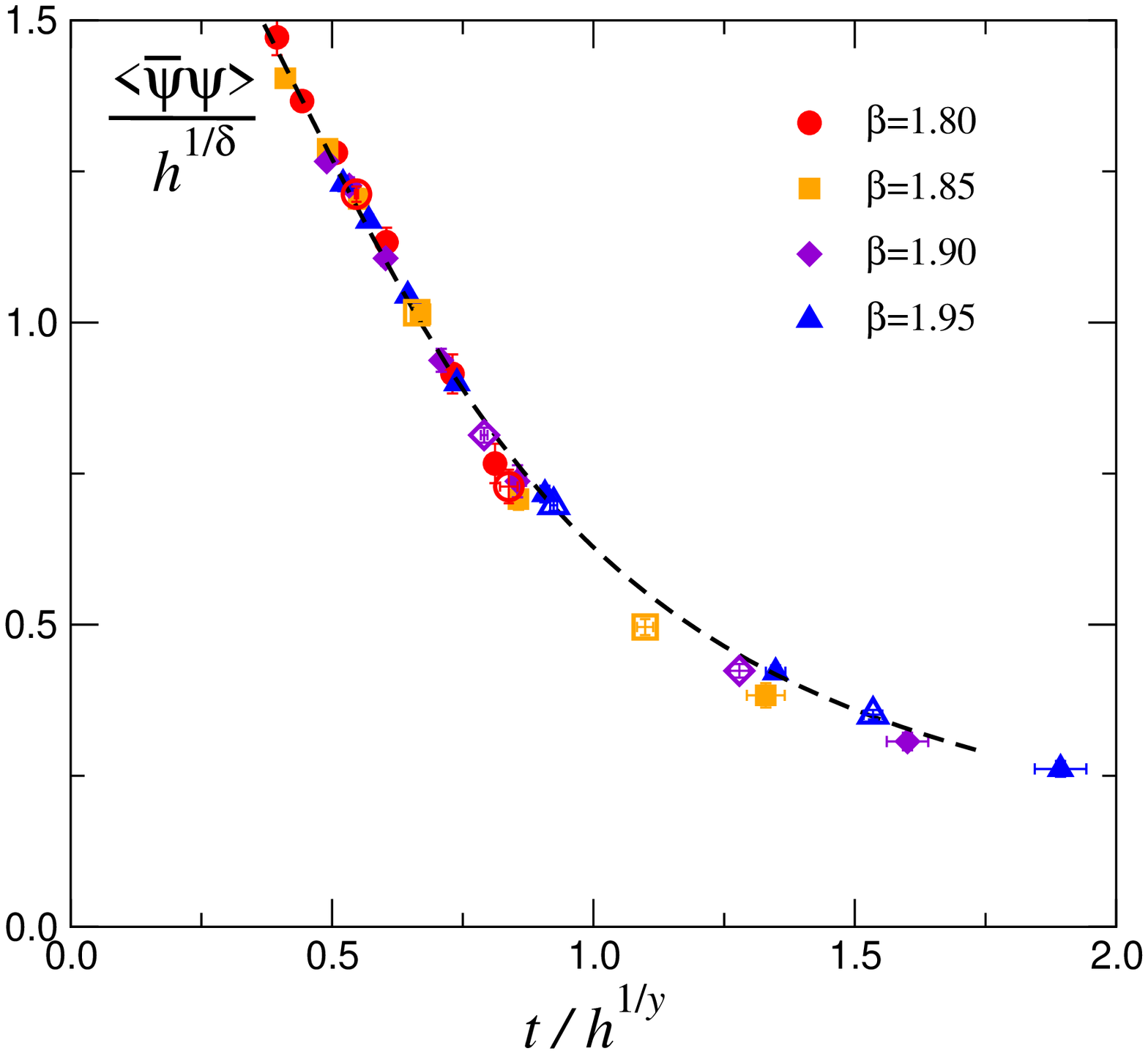}
\vskip -0.2cm
\caption{
Left: The ciral order parameter in 2-flavor QCD as a function of the Ward-identity quark mass for each $\beta$.
Right: O(4) scaling plot of the chiral order parameter. 
}
\label{fig2}
\end{center}
\vskip -0.3cm
\end{figure}

We plot the data of the chiral condensate as a function of the quark mass in the left panel of Fig.~\ref{fig2}. The closed symbols are obtained in the study of Ref.~\cite{cppacs1} by CP-PACS Collaboration, and the open symbols are new data in this study. (See Sec.~\ref{sec:simu}.)
We then reconstruct the data into $M/h^{1/\delta} = f(t/h^{1/y})$ for $\beta=1.80 - 1.95$ and $2m_q a = 0.0 - 0.9$, which are shown in Fig.~\ref{fig2} (right).
The critical exponents of the $O(4)$ spin model are used. 
The dashed line is the scaling function obtained by the $O(4)$ spin model in Ref.~\cite{toussaint}. We adjust three fit parameters in this analysis. One is the critical value of $\beta$, which is $\beta_{ct} = 1.462(66)$ in Ref.~\cite{cppacs1}, and the others are used for adjusting the scales of the horizontal and vertical-axes to the scaling function of the $O(4)$ spin model.
This scaling plot indicates that the scaling function of 2-flavor QCD is consistent with that of the $O(4)$ spin model.

\section{Scaling property at finite density}
\label{sec:dense}

Next, we discuss the universality property of 2-flavor QCD at finite density.
It is expected that, when $\mu_q$ is increased from 0 in the chiral limit, the second order phase transition changes to first order at the tricritical point, as shown in the right panel of Fig.~1.
The $O(4)$ scaling behavior is expected around the second order transition line even at finite densities because the action has the chiral symmetry in the chiral limit. 
At small $\mu_q$, we may identify the scaling variables, $t$ and $h$, by the following way for finite density QCD:
\begin{eqnarray}
M = \langle \bar{\psi} \psi \rangle,
\hspace{5mm}
t= \beta - \beta_{ct} + \frac{c}{2} \left(\frac{\mu_q}{T} \right)^2, 
\hspace{5mm}
h=2m_q a.
\end{eqnarray}
Here, $c$ is the curvature of the critical line in the $(\beta, \mu_q/T)$ plane.
On the other hand, $h$ does not have a $\mu_q$-dependent term at low density. 
Because the critical line is expected to run along the $m_q=0$ axis in the low density region of the $(m_q, \mu_q/T)$ plane, $h$ is zero at $m_q=0$.

It is worth discussing the phase structure investigating the scaling behavior of the chiral order parameter.
One of the easiest ways for confirming this scaling property is to calculate the second derivative of the chiral order parameter. 
We expect the following scaling properties:
\begin{eqnarray}
\left. \frac{d^2 M}{d(\mu_q/T)^2} \right|_{\mu_q=0} 
= c \left. \frac{dM}{dt} \right|_{\mu_q=0}, \hspace{5mm}
\left. \frac{dM/dt}{h^{1/\delta-1/y}} \right|_{\mu_q=0} 
= \left. \frac{df(x)}{dx} \right|_{x=t/h^{1/y}}.
\label{eq:der2sc}
\end{eqnarray}

We study the second derivative performing numerical simulations of 2-flavor QCD 
$(N_{\rm f}=2)$ at $\mu_q=0$. 
The derivative of the chiral order parameter is computed by the following way.
We define
\begin{eqnarray}
{\cal C}_n = (2K)^2 \frac{\partial^n {\rm tr} 
\left( D^{-1} \gamma_5 D^{-1} \gamma_5 \right)} {\partial (\mu_q a)^n}, 
\hspace{5mm}
{\cal Q}_n = N_{\rm f} \frac{\partial^n \ln \det D} 
{\partial (\mu_q a)^n}, 
\label{eq:basic}
\end{eqnarray}
\begin{eqnarray}
&& {\cal A}_1 = \left\langle {\cal Q}_1 \right\rangle, \hspace{4mm}
{\cal A}_2 = \left\langle {\cal Q}_2 \right\rangle 
+\left\langle {\cal Q}_1^2 \right\rangle, \hspace{4mm}
{\cal F}_0 = 
\left\langle {\cal C}_0 \right\rangle, \nonumber \\ &&
{\cal F}_1 = \left\langle {\cal C}_1 \right\rangle 
+ \left\langle {\cal C}_0 {\cal Q}_1 \right\rangle, \hspace{4mm}
{\cal F}_2 = \left\langle {\cal C}_2 \right\rangle 
+ 2 \left\langle {\cal C}_1 {\cal Q}_1 \right\rangle 
+ \left\langle {\cal C}_0 {\cal Q}_2 \right\rangle 
+ \left\langle {\cal C}_0 {\cal Q}_1^2 \right\rangle.
\end{eqnarray}
Then, the derivatives of the chiral condensate are given by
\begin{eqnarray} 
\left\langle \bar{\psi} \psi \right\rangle
\biggr|_{\mu_q =0}
&=& \frac{2m_q a}{N_s^3 N_t} {\cal F}_0 , \\
\frac{\partial \left\langle \bar{\psi} \psi \right\rangle
}{\partial (\mu_q/T)}  \biggr|_{\mu_q =0} 
&=& 
\frac{2m_qa}{N_s^3 N_t^2} 
\left( {\cal F}_1 - {\cal F}_0 {\cal A}_1 \right) =0 , \\
\frac{\partial^2 \left\langle \bar{\psi} \psi \right\rangle
}{\partial (\mu_q/T)^2}  \biggr|_{\mu=0} 
&=& 
\frac{2m_q a}{N_s^3 N_t^3} 
\left( {\cal F}_2 -2 {\cal F}_1 {\cal A}_1 - {\cal F}_0 {\cal A}_2 
+2 {\cal F}_0 {\cal A}_1^2 \right) 
= \frac{2m_q a}{N_s^3 N_t^3} 
\left( {\cal F}_2 - {\cal F}_0 {\cal A}_2 \right) ,
\end{eqnarray}
where we used the properties that
${\cal A}_n$ and ${\cal F}_n$ are zero for odd $n$'s at $\mu_q =0$.
The operators, ${\cal C}_n, {\cal Q}_n$, can be calculated by a random noise method.

\section{Numerical simulations for the calculation of the chiral condensate}
\label{sec:simu}

\begin{table}[tbp]
 \begin{center}
 \caption{Simulation parameters for $m_{\rm PS}/m_{\rm V}=0.65$ (left) 
 and $m_{\rm PS}/m_{\rm V}=0.80$ (right) \cite{whot07}.}
 \label{tab1}
 {\renewcommand{\arraystretch}{1.2} \tabcolsep = 3mm
 \begin{tabular}{|cccc|c|cccc|}
 \cline{1-4} \cline{6-9}
 \multicolumn{1}{|c}{$\beta$} &
 \multicolumn{1}{c} {$K$}    & 
 \multicolumn{1}{c} {$T/T_{pc}$} & 
 \multicolumn{1}{c|}{Traj.} & 
 \multicolumn{1}{c} {} & 
 \multicolumn{1}{|c}{$\beta$} &
 \multicolumn{1}{c} {$K$}    & 
 \multicolumn{1}{c} {$T/T_{pc}$} & 
 \multicolumn{1}{c|}{Traj.} \\
 \cline{1-4} \cline{6-9}
 1.80 & 0.145127 & 1.07(4)  & 5000 & & 1.80 & 0.141139 & 0.93(5)  & 6000 \\
 1.85 & 0.143502 & 1.18(4)  & 5000 & & 1.85 & 0.140070 & 0.99(5)  & 6000 \\
 1.90 & 0.141849 & 1.32(5)  & 5000 & & 1.90 & 0.138817 & 1.08(5)  & 6000 \\
 1.95 & 0.140472 & 1.48(5)  & 5000 & & 1.95 & 0.137716 & 1.20(6)  & 6000 \\
 \cline{1-4} \cline{6-9}
 \end{tabular}}
 \end{center}
\end{table}

\begin{figure}[t]
\begin{center}
\includegraphics[width=2.8in]{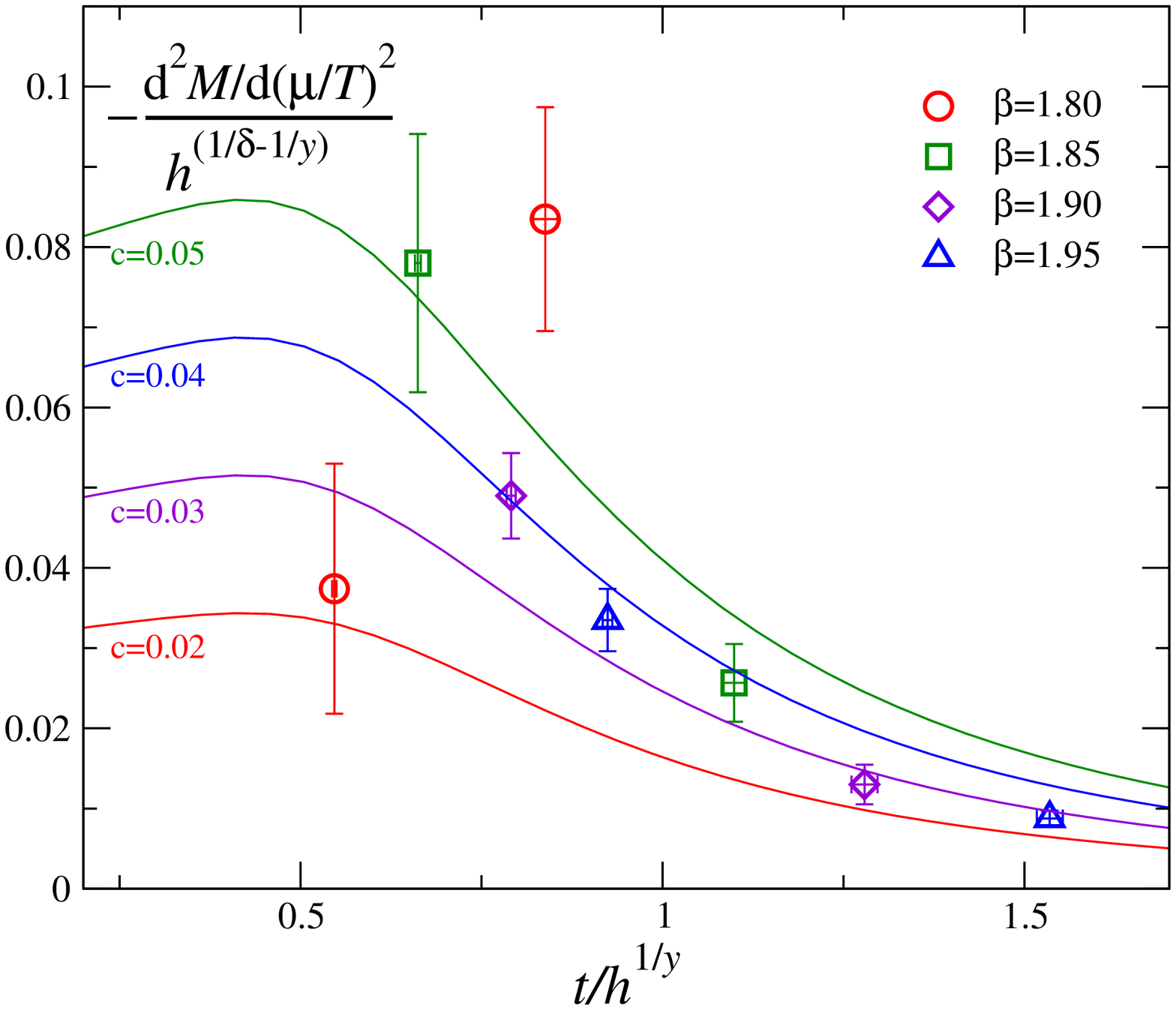}
\hskip 0.3cm
\includegraphics[width=2.8in]{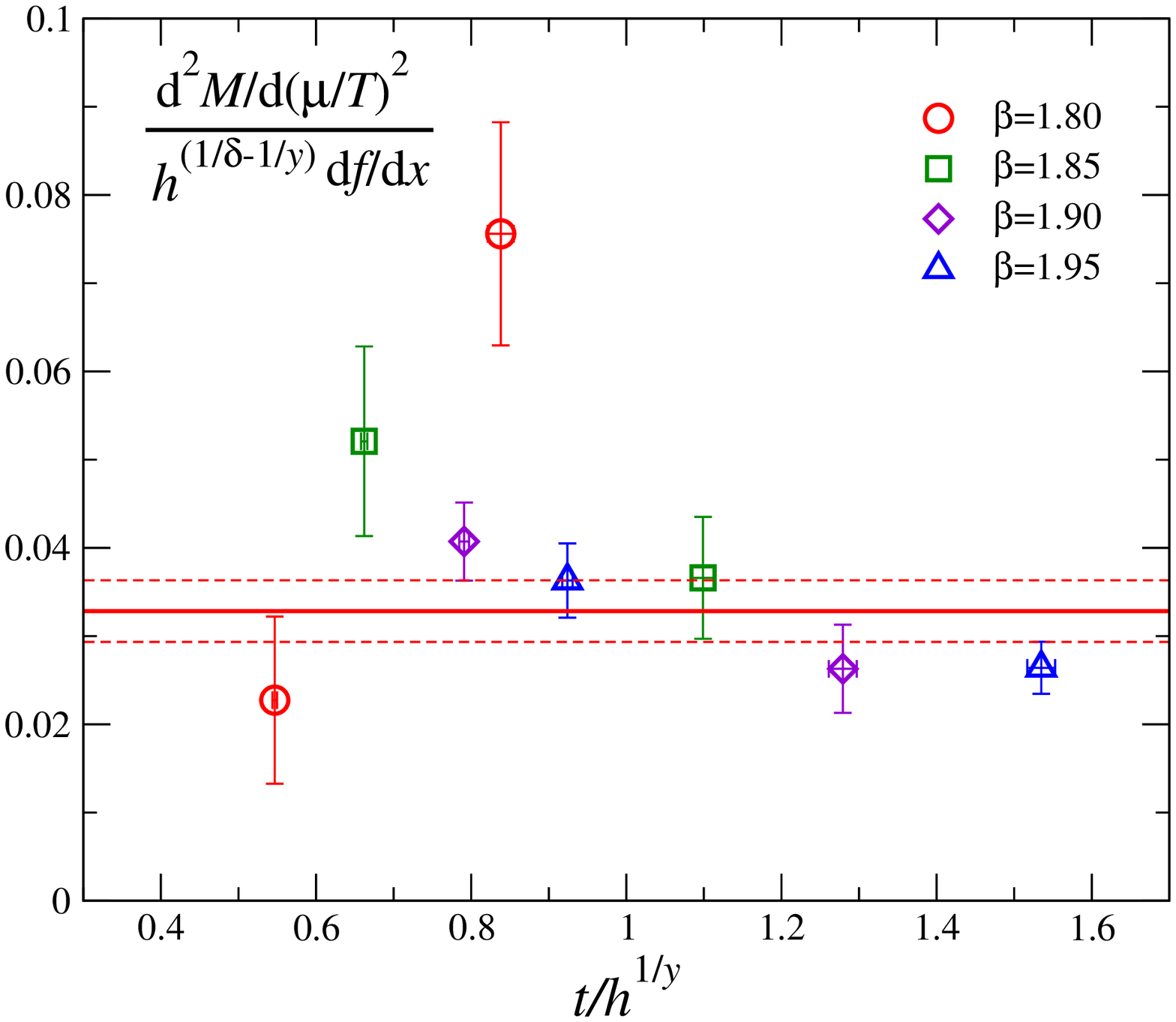}
\vskip -0.2cm
\caption{
Left: O(4) scaling plot of the second derivative of the chiral order parameter.
Right: The curvature of the critical line in the $(\beta, \mu_q/T)$ plane determined by 
the chiral order parameter. 
}
\label{fig3}
\end{center}
\vskip -0.3cm
\end{figure}

We calculate the second derivative of the chiral order parameter for 2-flavor QCD at $\mu_q=0$ using the configurations obtained in the simulations of Refs.~\cite{whot07,whot09}. 
The RG-improved gauge action and the $N_f=2$ clover-improved Wilson quark action are adopted. 
The simulation parameters are summarized in Table \ref{tab1}. 
The details of the simulation are given in Ref.~\cite{whot07,whot09}.

We used the random noise method of Ref.~\cite{whot09} for the calculation of the inverse of the matrices in Eq.~(\ref{eq:basic}). 
As we emphasized in Ref.~\cite{whot09}, it is important to apply the noise method only for the space index and to solve the inverse exactly for the spin and color indices without applying the noise method to obtain reliable results.
We choose the number of noise vectors 50 for each color and spin indices. 
The Ward-identity quark mass obtained by the zero temperature simulation in Ref.~\cite{cppacs1} is used for the scaling plot.  
It has been shown in Ref.~\cite{cppacs1} that the range $\beta > 1.95$ is out of the scaling region, i.e. the data of $\left\langle \bar{\psi} \psi \right\rangle$ do not agree with the $O(4)$ scaling function in that range. 
We thus plot the results of the second derivative of the chiral condensate at $\beta \le 1.95$ in the left panel of Fig.~\ref{fig3}. 
The colored line is the derivative of the scaling function $c \left( df(x)/dx \right)$, which is obtained by performing a numerical differentiation of $f(x)$ in the right panel of Fig.~\ref{fig2} numerically. 
The additional parameter $c$ is chosen as $c=0.02, 0.03, 0.04$ and $0.05$ from bottom to top.
These results are roughly consistent with the expected scaling behavior.
Because the error is still large, it is too early for drawing a definite conclusion. 
In particular, the error becomes large as $\beta$ decreases. 
But, the result suggests that the chiral order parameter satisfies the scaling relation in Eq.~(\ref{eq:der2sc}).

We then calculate the curvature of the critical line assuming the scaling relation. 
The ratio of $(d^2 M/d(\mu_q/T)^2) h^{-1/\delta +1/y}$ 
and $df(x)/dx|_{x=t/h^{1/y}}$ is plotted in Fig.~\ref{fig3} (right). 
This gives $c \equiv d^2 \beta_{ct}/d(\mu_q/T)^2$, i.e. the curvature of the critical line in the chiral limit.
Taking the average, we obtain $d^2 \beta_{ct}/d(\mu_q/T)^2 = -0.0328(35)$.
Moreover, the curvature of the critical temperature $T_c (\mu_q)$ at $\mu_q=0$ can be calculated by 
\begin{eqnarray}
\frac{1}{T_c}
\frac{d^2 T_c}{d(\mu_q/T)^2} 
= - \left. \frac{d^2 \beta_{ct}}{d(\mu_q/T)^2} \right/ a \frac{d \beta}{da}. 
\end{eqnarray}
In this calculation, we need $a(d\beta/da)$ at $\beta_{ct}$ in the chiral limit.
The value of $a(d\beta/da)$ for the lattice action which we used have been calculated in Ref.~\cite{cppacs2} for the pseudo-scalar-vector mass ratio $m_{PS}/m_{V} \geq 0.65$. 
If we try to estimate adopting the value near the pseudo-critical temperature at $m_{PS}/m_{V} = 0.65$, which is $a(d\beta/da) \approx -0.5$, the value of 
$(1/T_c) (d^2 T_c/d(\mu_q/T)^2) \approx -0.07$.
This value is smaller than that we expect from the data of the chemical freeze out, but this result is similar to the result in Ref.~\cite{bnl-bie10}. 
Further studies are, of course, necessary. However, it is found that this scaling analysis provides a systematic way to investigate the phase boundary of the finite density QCD in the low density region.

\section{Summary}
\label{sec:summary}

We studied the scaling property of the chiral phase transition in the low density region of finite temperature and density QCD. 
Assuming the phase structure shown in Fig.~\ref{fig1} (right), the scaling function in the low density region is discussed. 
To confirm the scaling property, we calculated the second derivative of the chiral order parameter performing a simulation of 2-flavor QCD with improved Wilson quarks and compared the results with an expected scaling function. 
Because the chiral symmetry is explicitly broken for the Wilson quark at finite lattice spacing, we used the chiral order parameter defined by a Ward-Takahashi identity for the scaling analysis. 
The scaling behavior of the chiral order parameter for 2-flavor QCD is roughly consistent with the scaling behavior of the corresponding spin model even at finite density.

Moreover, we calculated the curvature of the critical line in the chiral limit assuming the scaling relation. 
It is important to reduce the error by increasing the statistics and the number of noise vectors for more precise study.
However, we found that this scaling analysis provides a systematic way to investigate the phase boundary of finite density QCD in the low density region.

\vspace{5mm}
This work is in part supported 
by Grants-in-Aid of the Japanese Ministry of Education, Culture, Sports, Science and Technology, 
(Nos.  21340049, 22740168, 22840020, 20340047)
and by the Grant-in-Aid for Scientific Research on Innovative Areas
(Nos. 20105001, 20105003).

\end{document}